\documentclass[conference,10pt,letterpaper]{IEEEtran}
\usepackage[english]{babel}								
\makeatletter
\adddialect\l@ENGLISH\l@english
\makeatother

\usepackage[T1]{fontenc}

\usepackage{mathtools}
\usepackage{amsmath, amssymb, amsthm}
\usepackage[inline]{enumitem}

\usepackage{float}
\usepackage{stfloats}
\usepackage[section]{placeins}
\usepackage{balance}
\usepackage{cite}

\usepackage{xcolor}

\usepackage[pdftex]{graphicx}	
\usepackage[none]{hyphenat}

\newtheorem{theorem}{Theorem}

\newtheorem{proposition}[theorem]{Proposition}

\theoremstyle{remark}
\newtheorem{remark}{Remark}

\newtheoremstyle{colon}%
{}
{}
{\rm}
{}
{\itshape}
{:}
{ }
{\thmname{#1}\thmnumber{ \itshape#2}\thmnote{ (#3)}}

\theoremstyle{colon}

\newcommand{\dddag}{%
  \mathbin{\vbox{\offinterlineskip\ialign{%
    \hfil##\hfil\cr
    \small{$\dagger$}\cr
    \noalign{\kern-0.5ex}
    \small{$\ddagger$}\cr
}}}}

\begin{document}
\setlength{\abovedisplayskip}{3pt}
\setlength{\belowdisplayskip}{3pt}

\title{Rate-Splitting Meets Cell-Free MIMO Communications}

\author{\IEEEauthorblockN{André R. Flores$^{\dag,1}$, Rodrigo C. de Lamare$^{\dag,\ddag2}$ and Kumar Vijay Mishra$^{\dddag,3}$}
		\IEEEauthorblockA{$^\dag$Centre for Telecommunications Research, Pontifical Catholic University of Rio de Janeiro, Brazil}
		\IEEEauthorblockA{$^\ddag$Department of Electronic Engineering, University of York, UK}\vspace{-0.8em}
		\IEEEauthorblockA{$^{\dddag}$United States CCDC Army Research Laboratory, Adelphi, MD 20783 USA\\
			Email: 	\{$^1$andre.flores, $^2$delamare\}@cetuc.puc-rio.br,  $^3$kvm@ieee.org
			}
	}


%


\maketitle

\begin{abstract}
Multiuser multiple-input multiple-output wireless communications systems have the potential to satisfy the performance requirements of fifth-generation and future wireless networks. In this context, cell-free (CF) systems, where the antennas are distributed over the area of interest, have attracted attention because of their potential to enhance the overall efficiency and throughput performance when compared to traditional networks based on cells. However, the performance of CF systems is degraded by imperfect channel state information (CSI). To mitigate the detrimental effects of imperfect CSI, we employ rate splitting (RS) - a multiple-access scheme. The RS approach divides the messages of the users into two separate common and private portions so that interference is managed robustly. Unlike prior works, where the impact of RS in CF systems remains unexamined, we propose a CF architecture that employs RS with linear precoders to address deteriorated CSI. We derive closed-form expressions to compute the sum-rate performance of the proposed RS-CF architecture. Our numerical experiments show that our RS-CF system outperforms existing systems in terms of sum-rate, obtaining up to $10$\% higher gain. 
\end{abstract}


%
\IEEEpeerreviewmaketitle

\section{Introduction}
Future wireless networks are envisaged to provide new services such as augmented reality, virtual reality, and internet of things (IoT). These applications require higher data rates, denser connectivity and lower latency than traditional wireless systems. In this context, cell-free (CF) multiple-input multiple-output (MIMO) systems have emerged as a potential technology to satisfy these demands of modern applications \cite{Elhoushy2021}. The CF MIMO systems are shown to have 
higher energy efficiency (EE) and throughput per user than conventional cellular systems \cite{Ngo2017,Nguyen2017}. As a result, 
CF deployment has lately garnered 
significant research interest. 
These systems employ multiple distributed access points (APs) that are connected to a central processing unit (CPU). The APs serve a small group of users which are also geographically distributed. 

Similar to conventional MIMO systems \cite{Tse2005}, 
CF network also suffers from multiuser interference (MUI). 
To address MUI in the CF downlink, precoding techniques are employed. For instance, conjugate beamforming (CB) or matched filter (MF) precoder is a popular a low complexity precoding technique for CF \cite{Ngo2017}. In \cite{Nguyen2017}, zero-forcing (ZF) precoder for CF systems was investigated. The CB and ZF precoders were designed in \cite{Nayebi2017} with power allocation techniques to provide rate fairness among users. Recently, scalable minimum mean-square error (MMSE) precoders and combiners, which facilitate the deployment for CF systems, have also been proposed \cite{Bjoernson2020}. In \cite{Palhares2021a}, an iterative MMSE precoding was combined with a power allocation scheme.

In general, the precoder design assumes that the transmitter has perfect knowledge of the channel state information at the transmitter (CSIT) expressed as the channel matrix. It is obtained by employing pilot sequences along with reciprocity properties in systems using time division duplex (TDD) and feedback channels in frequency division duplex (FDD) systems \cite{Vu2007}. However, in practice, knowledge of perfect CSIT is an unrealistic assumption because channel estimate suffers from several sources of errors. For example, pilot contamination from noise adversely affects the estimation procedure. Moreover, time-varying dynamic channel requires constant CSIT update. Consequently, the transmitter has access to only partial or imperfect CSIT. Therefore, the precoder design is unable to mitigate the MUI leading to a residual MUI at the receiver. This residual MUI 
scales with the transmit power \cite{Tse2005}. Therefore, transmit techniques that can deal with imperfect CSIT are important. 


 Recently, rate-splitting (RS) \cite{Clerckx2016} has emerged as a novel approach to address CSIT imperfections more effectively than other conventional schemes. Indeed, RS outperforms conventional schemes such as the  precoding in spatial division multiplexing (SDMA) and the power-domain non-orthogonal multiple access (NOMA), by achieving higher rates than both schemes \cite{Mao2018}. Surprisingly, RS has shown benefits even against dirty paper coding \cite{Mao2020}. Interestingly, RS constitutes a generalized framework which has other transmission schemes such as SDMA, NOMA and multicasting as special cases \cite{Clerckx2020,Naser2020,Jaafar2020}.  The RS operates by splitting the message of one or several users into common and private messages. The common message must be decoded by all the users while the private messages are decoded only by their corresponding users. By adjusting the content and the power of the common message, RS controls how much interference should be decoded and how much interference is treated as noise, which is the major strength of this scheme. 
 
 Initially proposed in \cite{Han1981}, RS dealt with the interference channels \cite{Carleial1978}, where independent transmitters sent information to independent receivers \cite{Haghi2021}. This was extended in \cite{Yang2013} to the broadcast channel of a multiple-antenna system, where RS was shown to provide gains in terms of degrees-of-freedom (DoFs) with respect to conventional multi-user (MU) MIMO under imperfect CSIT. 
 In particular, \cite{Piovano2017} showed that RS achieves the optimal DoF region under imperfect CSIT. 
 
 Since then, several deployments and performance metrics involving RS were studied. The authors of \cite{JoudehClerckx2016,Hao2015} investigate the sum-rate maximization in multiple-input single-output (MISO) networks employing RS along with linear precoders. Not only the sum-rate maximization criterion has been considered but also a max-min fairness approach has been presented in \cite{Joudeh2017}, where overloaded multigroup multicasting scenarios where also considered. In \cite{Lu2018} RS was proposed for a system with random vector quantization feedback. In \cite{Flores2020,rsthp}, RS with common stream combining techniques was developed to exploit multiple antennas at the receiver and improve the overall sum-rate performance. In \cite{Li2020}, RS was employed in a single cell along with algorithms to reduce the number of streams and perform successive decoding for large number of users. However, the application of RS in CF systems remains relatively unexplored in previous studies. 
 

 In this paper, we propose a CF architecture that employs RS to send the information to the users. We perform singular value decomposition (SVD) over the channel matrix to obtain the common precoder. To send the private symbols, we employ MF and ZF precoders. Moreover, we derive expressions for signal-to-interference-plus-noise (SINR) and the ergodic-sum-rate (ESR) of the proposed RS-based CF system. Numerical experiments show that the RS strategy benefits from the  decentralized deployment of CF systems in terms of ESR performance. Contrary to prior works, where the ESR saturates quickly because of CSIT imperfections, we show that combining CF with RS yields a consistently increasing ESR even in the high SNR regime, robustness against imperfect CSIT, and significant sum-rate performance gains over standard CF and MU-MIMO networks. 

The rest of this paper is organized as follows. In Section II, we describe the system model of a CF MIMO system. We introduce the proposed RS-based CF system in Section~\ref{sec:rs} and present derivations of performance metrics in Section~\ref{sec:sr}. We validate our approaches through numerical examples in Section~\ref{sec:numexp} and give the conclusions in Section~\ref{sec:summ}.

Throughout the paper, we reserve bold lowercase, bold uppercase and calligraphic letters for the vectors, matrices, and sets respectively; $\textrm{Tr}(\cdot)$ and $\mathbb{E}\left[\cdot\right]$ represent trace and statistical expectation operators, respectively; the notations $(\cdot)^{\text{T}}$, $(\cdot)^H$, $(\cdot)^{*}$, $\lVert\cdot\rVert$ and $|\cdot|$ denote the transpose, Hermitian, complex conjugate, Euclidean norm, and magnitude, respectively. A diagonal matrix with the diagonal vector $\mathbf{v}$ is $\textrm{diag}(\mathbf{v})$. An $N \times K$ matrix with column vectors $\mathbf{a}_1$, $\cdots$, $\mathbf{a}_K$, each of length $N$, is $\mathbf{A} = \left[\mathbf{a}_1,\cdots,\mathbf{a}_K\right]$.  The operator $\Re\left\lbrace \cdot \right\rbrace$ retains the real part of a complex argument. The notation $a \sim \mathcal{CN}(0,\sigma_a^2)$ denotes circularly symmetric complex Gaussian random variable $a$ with zero mean and variance $\sigma_a^2$.

\section{System Model}
\label{sec:sysmod}
\begin{figure}[t]
\begin{center}
\includegraphics[width=0.8\columnwidth]{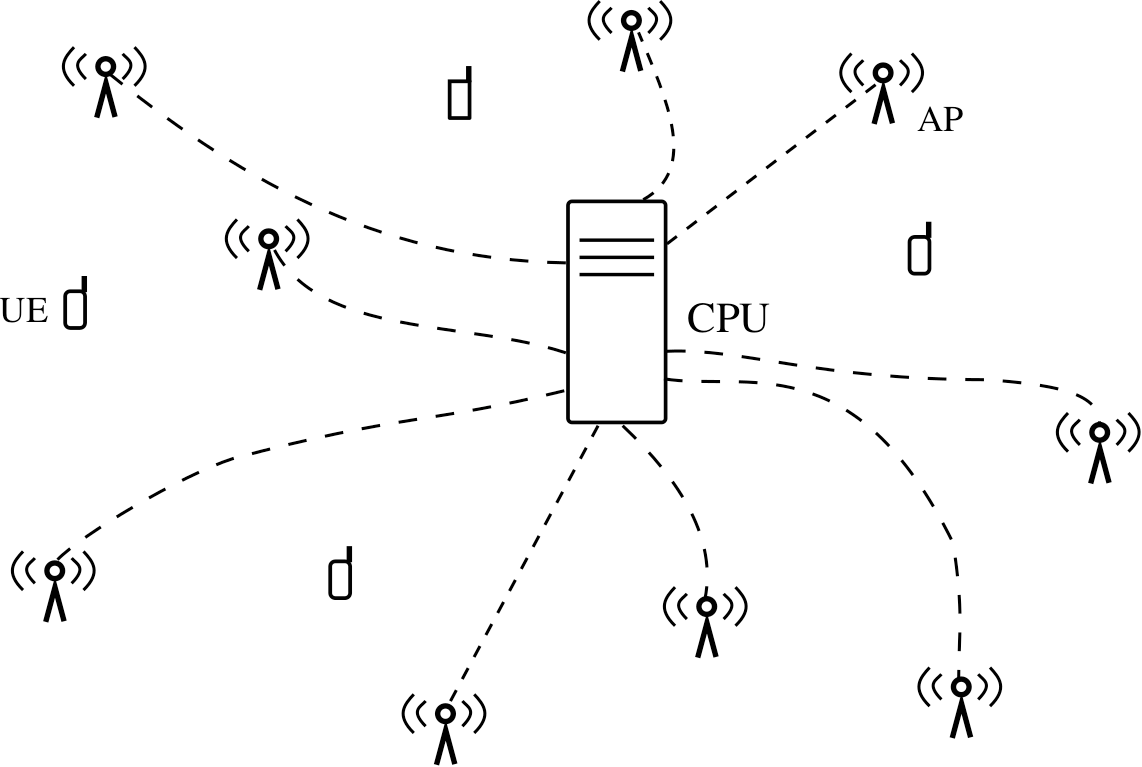}
\vspace{-1.0em}
\caption{Illustration of the CF MIMO system with randomly distributed APs and UEs. The CPU coordinates the APs.} 
\label{Fig1}
\end{center}
\end{figure}
Consider the downlink of a wireless network, in which the cells have not been partitioned. As a result, the users are not allocated to a specific base station (BS). The $M$ single antenna APs are distributed randomly and cover the geographical area of interest. Assume that $K$ users are randomly distributed in an under-loaded regime i.e., $M> K$. In this network, several antennas placed at different geographical points provide service to single-antenna user equipment (UE). This is different from a conventional cellular network that assigns the users to specific BSs. All antennas are connected to an RS-based central processing unit (CPU) that may be located in a cloud. 

The information is encoded into a vector of symbols $\mathbf{s}=\left[s_1,\cdots,s_K\right]^{\textrm{T}}\in \mathbb{C}^{K}$, where $s_k$ contains the data intended for the $k$-th user. The matrix $\mathbf{A}=\textrm{diag}\left(a_1,a_2,\cdots,a_K\right)\in\mathbb{R}^{K \times K}$ denotes the power allocation matrix which assigns a fraction of the available transmit power $P_t$ to each symbol. 
Then, a precoder $\mathbf{P}=\left[\mathbf{p}_1,\cdots,\mathbf{p}_K\right]\in \mathbb{C}^{M\times K}$ maps the symbols to the corresponding APs, which send the transmit vector $\mathbf{x}=\mathbf{P}\mathbf{A}\mathbf{s}$ to the users. The system follows a transmit power constraint, i.e., $\mathbb{E}\left[\lVert\mathbf{x}\rVert^2\right]\leq P_t$.

We employ a flat-fading channel model where the channel coefficient between $m$-th AP and $k$-th user is 
\begin{equation}
    g_{m,k}=\sqrt{\zeta_{m,k}}h_{m,k},
\end{equation}
where $\zeta_{m,k}$ is the large-scale fading coefficient that incorporates the path loss and shadowing effects 
and $h_{m,k}$ is an independently and identically distributed (i.i.d.) random variable that represents the small-scale fading coefficient and follows the 
distribution $\mathcal{CN}\left(0,1\right)$. 
Let us denote by $\tau_c$ the length in samples of the coherence time, which is the interval where the channel impulse response do not change. Then, the small-scale fading coefficients do not change 
during the coherence interval. 
The uplink channel between all APs and users is  $\mathbf{G}\in \mathbb{C}^{M\times K}$. 

The system employs TDD protocol and, therefore, the channel is estimated by exploiting the channel reciprocity property and pilot training. First, all users simultaneously and synchronously transmit the pilot sequences 
$\boldsymbol{\pi}_{1},\cdots,\boldsymbol{\pi}_{K}\in\mathbb{C}^{\tau}$. After receiving the pilots, the CPU computes the channel estimate $\mathbf{\hat{G}}=\left[\mathbf{\hat{g}}_1,\mathbf{\hat{g}}_2,\cdots,\mathbf{\hat{g}}_k\right]\in\mathbb{C}^{M\times K}$ and uses it for precoding. The error in the channel estimate is the ${M\times K}$ matrix $\mathbf{\tilde{G}}=\left[\mathbf{\tilde{g}}_1,\mathbf{\tilde{g}}_2,\cdots,\mathbf{\tilde{g}}_k\right]$, which is equal to $\mathbf{\tilde{G}}=\mathbf{G}-\hat{\mathbf{G}}$. The variable $\hat{g}_{m,k}$ denotes the channel estimate between the $m$th AP and the $k$th user and $\tilde{g}_{m,k}$ represents the error affecting the channel estimate $\hat{g}_{m,k}$. Specifically, 
\begin{align}
    \hat{g}_{m,k}\sim\mathcal{CN}\left(0,\left(1-\sigma^2_{e}\right)\zeta_{m,k}\right), && \tilde{g}_{m,k}\sim\mathcal{CN}\left(0,\sigma^2_{e}\zeta_{m,k}\right),
\end{align}
where $\sigma^2_{e}$ models different levels of imperfect CSI. 
The received signal is 
\begin{equation}
    \mathbf{y}=\mathbf{G}^{\text{T}}\mathbf{P}\mathbf{A}\mathbf{s}+\mathbf{w},
\end{equation}
where the vector $\mathbf{w}=\left[w_1,w_2,\cdots,w_K\right]^{\text{T}}\sim\mathcal{CN}\left(\mathbf{0},\sigma_w\mathbf{I}\right)$ represents the additive white Gaussian noise (AWGN) at the receiver. 

\section{RS-Based CF Deployment}
\label{sec:rs}
In our proposed RS-based CF, both common and private messages are encoded and modulated into a vector of symbols. The common symbol $s_c$ is superimposed on the private symbols $\mathbf{s}$ to yield the symbol vector $\mathbf{s}^{\left(\textrm{RS}\right)}=\left[s_c,s_1,\cdots,s_K\right]^{\textrm{T}}\in \mathbb{C}^{K+1}$.  
The power allocation matrix is $\mathbf{A}^{\left(\text{RS}\right)}=\text{diag}\left(a_c,a_1,a_2,\cdots,a_K\right)$, where $a_c$ is the coefficient that allocates a fraction of the available power $P_t$ to the common symbol, wheres $a_i, i=1,\ldots, K$ are the coefficients that allocate the powers to the private streams. To map the common symbol to the transmit antennas, a common precoder $\mathbf{p}_c$ is included into the precoding matrix $\mathbf{P}^{\left(\textrm{RS}\right)}=\left[\mathbf{p}_c,\mathbf{p}_1,\cdots,\mathbf{p}_K\right]\in \mathbb{C}^{M\times\left(K+1\right)}$. The APs send the transmit vector $\mathbf{x}^{\left(\text{RS}\right)}=\mathbf{P}^{\left(\textrm{RS}\right)}\mathbf{A}^{\left(\text{RS}\right)}\mathbf{s}^{\left(\textrm{RS}\right)}$ with the private and common data to the users. The received signal at the $k$-th user is\par\noindent\small
\begin{align}
    y_k=a_c s_c \mathbf{\hat{g}}_{k}^{\textrm{T}}\mathbf{p}_c+\underbrace{\sum_{i=1}^Ka_is_i\mathbf{\hat{g}}_k^{\textrm{T}}\mathbf{p}_i}_{\textrm{MUI}}+\underbrace{a_c s_c \mathbf{\tilde{g}}_{k}^{\textrm{T}}\mathbf{p}_c+\sum_{j=1}^K a_j s_j\mathbf{\tilde{g}}_k^{\textrm{T}}\mathbf{p}_j}_{\textrm{Residual Interference}}+w_k. \label{received signal user k}
\end{align}\normalsize
The vector with information of all users is\par\noindent\small
\begin{align}
    \mathbf{y}=&a_c s_c \mathbf{\hat{G}}^{\textrm{T}}\mathbf{p}_c+\sum_{i=1}^K a_i s_i \mathbf{\hat{G}}^{\textrm{T}}\mathbf{p}_i+a_c s_c\mathbf{\tilde{G}}^{\textrm{T}}\mathbf{p}_c+\sum_{j=1}^K a_j s_j \mathbf{\tilde{G}}^{\textrm{T}}\mathbf{p}_j+\mathbf{w}.
\end{align}\normalsize

At the receiver, first the common message is decoded and then successive interference cancellation (SIC) is performed to subtract the common information from the received signal. Note that we assume that perfect CSIT is available at the receiver. Removing the common symbol at the $k$-th user yields\par\noindent\small
\begin{equation}
    y_k=a_k s_k \mathbf{\hat{g}}_k^{\textrm{T}}\mathbf{p}_k+\sum_{\substack{i=1\\i\neq k}}^K a_i s_i\mathbf{\hat{g}}_k^{\textrm{T}}\mathbf{p}_i+\sum_{j=1}^K a_j s_j\mathbf{\tilde{g}}_k^{\textrm{T}}\mathbf{p}_j+w_k.
\end{equation}\normalsize
Thereafter, the receiver recovers its private message. In what follows, we present the design of MF and ZF precoders for the proposed RS-based CF system, which are simple and widely-used precoding techniques.

\subsection{MF precoder design}
We employ a MF precoder to send the private symbols due to its reduced computational complexity. 
This precoder is simply obtained by the Hermitian of the channel estimate \cite{Joham2005}, i.e., $\mathbf{P}=\mathbf{\hat{G}}^*$. 
To obtain the common precoder, we perform SVD over the channel estimate \cite{Clerckx2016,Flores2020}. This produces $\mathbf{\hat{G}}^{\text{T}}=\mathbf{U}\mathbf{\Psi}\mathbf{V}^{H}$, where $\mathbf{\Psi}\in\mathbb{R}^{K \times M}$ is a matrix that has zeros outside its main diagonal. The $k$th element in the main diagonal of $\mathbf{\Psi}$ is denoted by $\psi_k$. Additionally, the matrices $\mathbf{U}=\left[\mathbf{u}_{1,*}^{\text{T}},\mathbf{u}_{2,*}^{\text{T}},\cdots,\mathbf{u}_{K,*}^{\text{T}}\right]^{\text{T}} \in \mathbb{C}^{\left(K\times K\right)}$ and $\mathbf{V}=\left[\mathbf{v}_1,\mathbf{v}_2,\cdots,\mathbf{v}_K\right]\in \mathbb{C}^{M\times M}$ are unitary matrices where $\mathbf{u}_{k,*}$ denotes the $k$-th row of matrix $\mathbf{U}$. Then, we set the common precoder equal to the first column of matrix $\mathbf{V}$, i.e., $\mathbf{p}_c=\mathbf{v}_1$. Then, the precoder employed in the proposed architecture is \par\noindent\small
\begin{equation}
    \mathbf{P}^{\left(\text{RS}\right)}=\left[\mathbf{v}_1,\mathbf{\hat{g}}_1^*,\mathbf{\hat{g}}_2^*,\cdots,\mathbf{\hat{g}}_K^*\right].\label{MF plus SVD}
\end{equation}\normalsize
Then, the received signal at the $k$-th user is\par\noindent\small
\begin{align}
  y_k=a_c s_c\mathbf{\hat{g}}^{\text{T}}_k\mathbf{v}_1+\sum_{i=1}^K a_is_i\mathbf{\hat{g}}_k^{\textrm{T}}\mathbf{\hat{g}}^*_i+a_c s_c \mathbf{\tilde{g}}_{k}^{\textrm{T}}\mathbf{v}_1+\sum_{j=1}^K a_j s_j\mathbf{\tilde{g}}_k^{\textrm{T}}\mathbf{\hat{g}}^*_j+w_k. 
\end{align}\normalsize

\subsection{ZF precoder design}
The ZF precoder is the pseudoinverse of the channel estimate \cite{Joham2005}, i.e., \par\noindent\small
\begin{equation}
\mathbf{P}=\mathbf{\hat{G}}^{*}\left(\mathbf{\hat{G}}^{\text{T}}\mathbf{\hat{G}}^{*}\right)^{-1}.
\end{equation} \normalsize
Let us introduce the matrix $\boldsymbol{\Lambda}=\left[\boldsymbol{\lambda}_1,\boldsymbol{\lambda}_2,\cdots,\boldsymbol{\lambda}_K\right] $. Considering that $\boldsymbol{\Lambda}=\left(\mathbf{\hat{G}}^{\text{T}}\mathbf{\hat{G}}^{*}\right)^{-1}$ gives $\mathbf{P}^{\left(\text{RS}\right)}=\left[\mathbf{v}_1,\mathbf{\hat{G}}^{*}\boldsymbol{\lambda}_1,\cdots,\mathbf{\hat{G}}^{*}\boldsymbol{\lambda}_K\right]$. The received signal is\par\noindent\small
\begin{align}
     y_k=&a_c s_c\mathbf{\hat{g}}^{\text{T}}_k\mathbf{v}_1+ a_k s_k+a_c s_c \mathbf{\tilde{g}}_{k}^{\textrm{T}}\mathbf{v}_1
    +\sum_{i=1}^K a_i s_i\mathbf{\tilde{g}}_k^{\textrm{T}}\mathbf{\hat{G}}^{*}\boldsymbol{\lambda}_i+w_k. 
\end{align}\normalsize

\subsection{Power allocation}
The proposed precoders must fulfil the transmit power constraint, i.e., $\mathbb{E}\left[\lVert\mathbf{x}^{\left(\text{RS}\right)}\rVert^2\right]\leq P_t$. Note that a part of the power must be reserved to transmit the common symbol. Given $a_c^2=\delta P_t$, 
where $\delta \in \left[0,1\right]$ represents the fraction of power allocated to the common stream. Then, the power available for transmitting the private streams is $\left(1-\delta\right)P_t$. The performance of the RS structure depends on an appropriate power allocation, which in this case translates to find an appropriate value of $\delta$. If $\delta$ is properly set, the performance of RS-CF should be at least as good as the performance obtained by CF without RS because the system can set $\delta=0$ implying that the splitting procedure is avoided.

In this work, we employ an exhaustive search to find a suitable value $\delta^{\left(o\right)}$ given that uniform power allocation is used across the private symbols. The value $\delta^{\left(o\right)}$ obeys\par\noindent\small
\begin{equation}
    \delta^{\left(o\right)}= \max_{\delta}S_r\left(\delta\right),
\end{equation}\normalsize
where $\max_{\delta}S_r\left(\delta\right)$ denotes the ESR and is explained in more detail in Section IV.
Other approaches to perform power allocation in RS are also of interest, such as convex optimization \cite{JoudehClerckx2016} and monotonic optimization \cite{Tuy2000} techniques.

\section{Sum-Rate performance}
\label{sec:sr}
From \eqref{received signal user k}, average power of received signal at $k$-th user is\par\noindent\small
\begin{equation}
    \mathbb{E}\left[\lvert y_k\rvert^2\right]=a_c^2\lvert \mathbf{g}_k^{\textrm{T}}\mathbf{p}_c\rvert^2+\sum_{i=1}^K a_i^2\lvert \mathbf{g}_k^{\textrm{T}}\mathbf{p}_i\rvert^2+\sigma_w^2.
\end{equation}\normalsize
While decoding the common symbol at the $k$-th user, the instantaneous SINR is\par\noindent\small
\begin{align}
    \gamma_{c,k}&=\frac{a_c^2\lvert \mathbf{\hat{g}}_k^{\textrm{T}}\mathbf{p}_c\rvert^2}{d_{c,k}+\sum\limits_{i=1}^K a_i^2\lvert \mathbf{g}_k^{\textrm{T}}\mathbf{p}_i\rvert^2+\sigma_w^2},\label{instantaneous SINR common rate}
\end{align}\normalsize
where $d_{c,k}=2\Re\left\{\left(\mathbf{\hat{g}}_k^{\text{T}}\mathbf{p}_c\right)^*\left(\mathbf{\tilde{g}}_k^{\text{T}}\mathbf{p}_c\right)\right\}+a_c^2\lvert\mathbf{\tilde{g}}^{\text{T}}_k\mathbf{p}_c\rvert^2$ corresponds to the power loss arising from the error in the channel estimate. After applying SIC, the instantaneous SINR while decoding the private symbol at the $k$-th user is \par\noindent\small
\begin{equation}
    \gamma_k=\frac{a_k^2\lvert\mathbf{\hat{g}}_k^{\textrm{T}}\mathbf{p}_k\rvert^2}{d_k+\sum\limits_{\substack{i=1\\i\neq k}}^K a_i^2\lvert\mathbf{g}_k\mathbf{p}_i\rvert^2+\sigma_w^2},\label{instantaneous SINR private rate}
\end{equation}\normalsize
where $d_{k}=2\Re\left\{\left(\mathbf{\hat{g}}_k^{\text{T}}\mathbf{p}_k\right)^*\left(\mathbf{\tilde{g}}_k^{\text{T}}\mathbf{p}_k\right)\right\}+a_c^2\lvert\mathbf{\tilde{g}}^{\text{T}}_k\mathbf{p}_k\rvert^2$. 

For Gaussian signaling, the instantaneous common rate (CR) at the $k$-th user is 
\begin{equation}
    R_{c,k}=\log_2\left(1+\gamma_{c,k}\right).\label{instantaneous common rate per user}
\end{equation}
The private rate at user $k$ is 
\begin{equation}
     R_{k}=\log_2\left(1+\gamma_{k}\right)\label{instantaneous private rate}
\end{equation}
The instantaneous rates are not achievable if we consider imperfect CSIT. Therefore, we employ average sum-rate (ASR) 
wherein the effects of errors in the channel estimate are averaged out. The ASR consists of two parts, the average common rate (CR) and the average private rate. Following this definition, the average CR at the $k$-th user is $\bar{R}_{c,k}=\mathbb{E}\left[R_{c,k}|\mathbf{\hat{G}}\right]$. Analogously, the average private rate at the $k$-th user is  $\bar{R}_{k}=\mathbb{E}\left[R_{k}|\mathbf{\hat{G}}\right]$. The performance of the system over a large number of channel realizations is measured by the ESR 
\begin{equation}
     S_r=\min_{k}\mathbb{E}\left[\bar{R_{c,k}}\right]+\sum_{k=1}^K \mathbb{E}\left[\bar{R}_k\right],\label{system ergodic sum rate}
 \end{equation}
 Note that in \eqref{system ergodic sum rate} we employ the minimum CR found across users in order to guarantee that all users decode the common symbol successfully.
 
 \begin{remark}
 Under perfect CSIT, the instantaneous sum rate is achievable. In such cases, $\mathbf{\tilde{G}}=\mathbf{0}$. The SINRs in \eqref{instantaneous SINR common rate} and \eqref{instantaneous SINR private rate} becomes, respectively,\par\noindent\small
 \begin{align}
\gamma_{c,k}&=\frac{a_c^2\lvert \mathbf{g}_k^{\textrm{T}}\mathbf{p}_c\rvert^2}{\sum\limits_{i=1}^K a_i^2\lvert \mathbf{g}_k^{\textrm{T}}\mathbf{p}_i\rvert^2+\sigma_w^2},\end{align}\normalsize
and\par\noindent\small
\begin{align}
\gamma_k&=\frac{a_k^2\lvert\mathbf{\hat{g}}_k^{\textrm{T}}\mathbf{p}_k\rvert^2}{d_k+\sum\limits_{\substack{i=1\\i\neq k}}^K a_i^2\lvert\mathbf{g}_k\mathbf{p}_i\rvert^2+\sigma_w^2}.
 \end{align}\normalsize
 The ESR is 
 \begin{equation}
 S_r=\min_{k}\mathbb{E}\left[R_{c,k}\right]+\sum_{k=1}^K \mathbb{E}\left[R_k\right],
 \end{equation}
 \end{remark}
 
 \subsection{Sum-rate of MF precoder}
To obtain ESR for the MF precoder, we derive a closed-form expression for the SINR in the following Proposition~\ref{prop:sinr_MF}.
\begin{proposition}
\label{prop:sinr_MF}
When decoding the common stream, the SINR is\par\noindent\small
\begin{equation}
     \gamma_{c,k}=\frac{a_c^2\psi_1^2\lvert u_{k,1}\rvert^2}{d_{c,k}^{\left(\text{v}\right)}+\sum_{i=1}^K a_i^2\lvert\left(\mathbf{\hat{g}}_k^{\text{T}}+\mathbf{\tilde{g}}^{\text{T}}_k\right)\mathbf{\hat{g}}_i^*\rvert^2+\sigma_w^2},\label{SINR when decoding common message MF approach}
 \end{equation}\normalsize
  with $d_{c,k}^{\left(\text{v}\right)}=a_c^2\left(2\psi_1\Re\left\{u_{k,1}^*\left(\mathbf{\tilde{g}}_k^{\text{T}}\mathbf{v}_1\right)\right\}+\lvert\mathbf{\tilde{g}}^{\text{T}}_k\mathbf{v}_1\rvert^2\right)$. When decoding the private message after SIC, the SINR is\par\noindent\small
    \begin{equation}
     \gamma_k=\frac{a_k^2\lVert\mathbf{\hat{g}}_k^{\text{T}}\rVert^4}{d_{k}^{\left(\text{MF}\right)}+\sum\limits_{\substack{i=1\\i\neq k}}^K a_i^2\lvert\left(\mathbf{\hat{g}}_k^{\text{T}}+\mathbf{\tilde{g}}^{\text{T}}_k\right)\mathbf{\tilde{g}}_i^*\rvert^2+\sigma_w^2},\label{SINR when decoding private message MF approach}
 \end{equation}\normalsize
   where $d_{k}^{\left(\text{MF}\right)}=a_k^2\left( 2\lVert\mathbf{\hat{g}}_k^{\text{T}}\rVert^2\Re\left\{\mathbf{\tilde{g}}_k^{\text{T}}\mathbf{\hat{g}}^*_k\right\}+\lvert\mathbf{\tilde{g}}_k^{\text{T}}\mathbf{\hat{g}}^*_k\rvert^2\right)$.
 \end{proposition}
 \begin{IEEEproof}
 From the precoder in \eqref{MF plus SVD}, we obtain the average power of the received signal at user $k$ as \par\noindent\small
 \begin{align}
     \mathbb{E}\left[\lvert y_k\rvert^2\right]=a_c^2\underbrace{\lvert\left(\mathbf{\hat{g}}_k^{\text{T}}+\mathbf{\tilde{g}}^{\text{T}}_k\right)\mathbf{v}_1\rvert^2}_{T_1}+\sum_{i=1}^K a_i^2\underbrace{\lvert\left(\mathbf{\hat{g}}_k^{\text{T}}+\mathbf{\tilde{g}}^{\text{T}}_k\right)\mathbf{\hat{g}}_i^*\rvert^2}_{T_2}+\sigma_w^2.\label{average receive power MF plus SVD}
 \end{align}\normalsize
 Expanding $T_1$ gives\par\noindent\small
 \begin{align}
     T_1=&\left(\mathbf{\hat{g}}^{\text{T}}_k\mathbf{v}_1+\mathbf{\tilde{g}}^{\text{T}}_k\mathbf{v}_1\right)^*\left(\mathbf{\hat{g}}^{\text{T}}_k\mathbf{v}_1+\mathbf{\tilde{g}}^{\text{T}}_k\mathbf{v}_1\right)\nonumber\\
     =&\lvert\mathbf{\hat{g}}^{\text{T}}_k\mathbf{v}_1\rvert^2+2\Re\left\{\left(\mathbf{\hat{g}}^{\text{T}}_k\mathbf{v}_1\right)^*\left(\mathbf{\tilde{g}}^{\text{T}}_k\mathbf{v}_1\right)\right\}+\lvert\mathbf{\tilde{g}}^{\text{T}}_k\mathbf{v}_1\rvert^2.\label{T1 evaluation}
 \end{align}\normalsize
 Observe that\par\noindent\small
 \begin{align}
     \mathbf{\hat{g}}^{\text{T}}_k\mathbf{v}_1= \mathbf{u}_{k,*}\boldsymbol{\Psi}\mathbf{V}^H\mathbf{v}_1
     =\mathbf{u}_{k,*},\boldsymbol{\psi}_1
     =u_{k,1}\psi_1.
 \end{align}\normalsize
Then, \eqref{T1 evaluation} becomes\par\noindent\small
\begin{equation}
    T_1=\psi_1^2\lvert u_{k,1}\rvert^2+2\psi_1\Re\left\{u_{k,1}^*\left(\mathbf{\tilde{g}}_k^{\text{T}}\mathbf{v}_1\right)\right\}+\lvert\mathbf{\tilde{g}}^{\text{T}}_k\mathbf{v}_1\rvert^2.\label{T1 reduced}
\end{equation}\normalsize

Expanding $T_2$ 
yields\par\noindent\small
\begin{align}
    T_2=&\left(\mathbf{\hat{g}}^{\text{T}}_k\mathbf{\hat{g}}_i^*+\mathbf{\tilde{g}}^{\text{T}}_k\mathbf{\hat{g}}_i^*\right)^*\left(\mathbf{\hat{g}}^{\text{T}}_k\mathbf{\hat{g}}_i^*+\mathbf{\tilde{g}}^{\text{T}}_k\mathbf{\hat{g}}_i^*\right)\nonumber\\
    =&\lvert\mathbf{\hat{g}}^{\text{T}}_k\mathbf{\hat{g}}^*_i\rvert^2+2\Re\left\{\left(\mathbf{\hat{g}}^{\text{T}}_k\mathbf{\hat{g}}_i^*\right)^*\left(\mathbf{\tilde{g}}^{\text{T}}_k\mathbf{\hat{g}}_i^*\right)\right\}+\lvert\mathbf{\tilde{g}}^{\text{T}}_k\mathbf{\hat{g}}_i^*\rvert^2,\label{T2 expanded}
\end{align}\normalsize
reducing \eqref{T2 expanded}, when $k=i$, to\par\noindent\small
\begin{equation}
    T_2=\lVert\mathbf{\hat{g}}_k^{\text{T}}\rVert^4+2\lVert\mathbf{\hat{g}}_k^{\text{T}}\rVert^2\Re\left\{\mathbf{\tilde{g}}_k^{\text{T}}\mathbf{\hat{g}}^*_k\right\}+\lvert\mathbf{\tilde{g}}_k^{\text{T}}\mathbf{\hat{g}}^*_k\rvert^2.\label{T2 reduced}
\end{equation}\normalsize
 
 By substituting \eqref{T1 reduced} and \eqref{T2 reduced} in \eqref{instantaneous SINR common rate} and \eqref{instantaneous SINR private rate} we get \eqref{SINR when decoding common message MF approach} and \eqref{SINR when decoding private message MF approach}.
 
\end{IEEEproof}
The ESR of the proposed system is obtained by substituting \eqref{SINR when decoding common message MF approach} and \eqref{SINR when decoding private message MF approach} into \eqref{instantaneous common rate per user}, \eqref{instantaneous private rate}, and \eqref{system ergodic sum rate}. 
 
 \begin{remark}
 Under perfect CSIT, the SINRs simplify as\par\noindent\small
 \begin{equation}
     \gamma_{c,k}=\frac{a_c^2\psi_1^2\lvert u_{k,1}\rvert^2}{\sum\limits_{i=1}^K a_i^2\lvert\mathbf{\hat{g}}_k^{\text{T}}\mathbf{\hat{g}}_i^*\rvert^2+\sigma_w^2},
 \end{equation}\normalsize
 and\par\noindent\small
 \begin{equation}
     \gamma_k=\frac{a_k^2\lVert\mathbf{\hat{g}}^{\text{T}}_k\rVert^4}{\sum\limits_{\substack{i=1\\i\neq k}}^K a_i^2\lvert\mathbf{\hat{g}}_k^{\text{T}}\mathbf{\hat{g}}_i^*\rvert^2+\sigma_w^2}.
 \end{equation}\normalsize
 \end{remark}
 
 \subsection{Sum-rate of ZF precoder}
Similar to the MF precoder, we have the following Proposition~\ref{prop:sinr_ZF} for the SINR of the ZF precoder.
 \begin{proposition}
 \label{prop:sinr_ZF}
  For the ZF, the SINR when decoding the common message is\par\noindent\small
 \begin{equation}
     \gamma_{c,k}=\frac{a_c^2\psi_1^2\lvert u_{k,1}\rvert^2}{d_k^{\left(v\right)}+a_k^2 +\sum_{i=1}^{K}a_i^2\lvert \mathbf{\tilde{g}}_k^{\text{T}}\mathbf{\hat{G}}^{*}\boldsymbol{\lambda}_i\rvert^2+\sigma_w^2},\label{SINR when decoding common message ZF approach}
 \end{equation}\normalsize
whereas the SINR when decoding the private message at $k$-th user is \par\noindent\small
\begin{equation}
    \gamma_k=\frac{a_k^2}{\lvert\mathbf{\tilde{g}}_k^{\text{T}}\mathbf{\hat{G}}^{*}\boldsymbol{\lambda}_i\rvert^2+\sigma_w^2}.\label{SINR when decoding private message ZF approach}
\end{equation}\normalsize
 \end{proposition}
 \begin{IEEEproof}
  For the ZF precoder, we have $\mathbf{\hat{g}}_k^{\text{T}}\mathbf{p}_k=1$ and $\mathbf{\hat{g}}_k^{\text{T}}\mathbf{p}_i=0$ for all $i\neq k$. Then, the average power of the received signal is\par\noindent\small
 \begin{equation}
     \mathbb{E}\left[\lvert y_k\rvert^2\right]=a_c^2 T_1+a_k^2 +\sum_{i=1}^{K}a_i^2\lvert \mathbf{\tilde{g}}_k^{\text{T}}\mathbf{\hat{G}}^{*}\boldsymbol{\lambda}_i\rvert^2+\sigma_w^2\label{Average power of the received signal ZF approach}
 \end{equation}\normalsize
 From \eqref{Average power of the received signal ZF approach} we can obtain \eqref{SINR when decoding common message ZF approach} and \eqref{SINR when decoding private message ZF approach}.
 \end{IEEEproof}

Similar to the MF case, the ESR is computed by substituting \eqref{SINR when decoding common message ZF approach} and \eqref{SINR when decoding private message ZF approach} into \eqref{instantaneous common rate per user}, \eqref{instantaneous private rate}, and \eqref{system ergodic sum rate}.

 \section{Numerical Experiments}
 \label{sec:numexp}
\begin{figure}[t]
\begin{center}
\includegraphics[width=1.0\columnwidth]{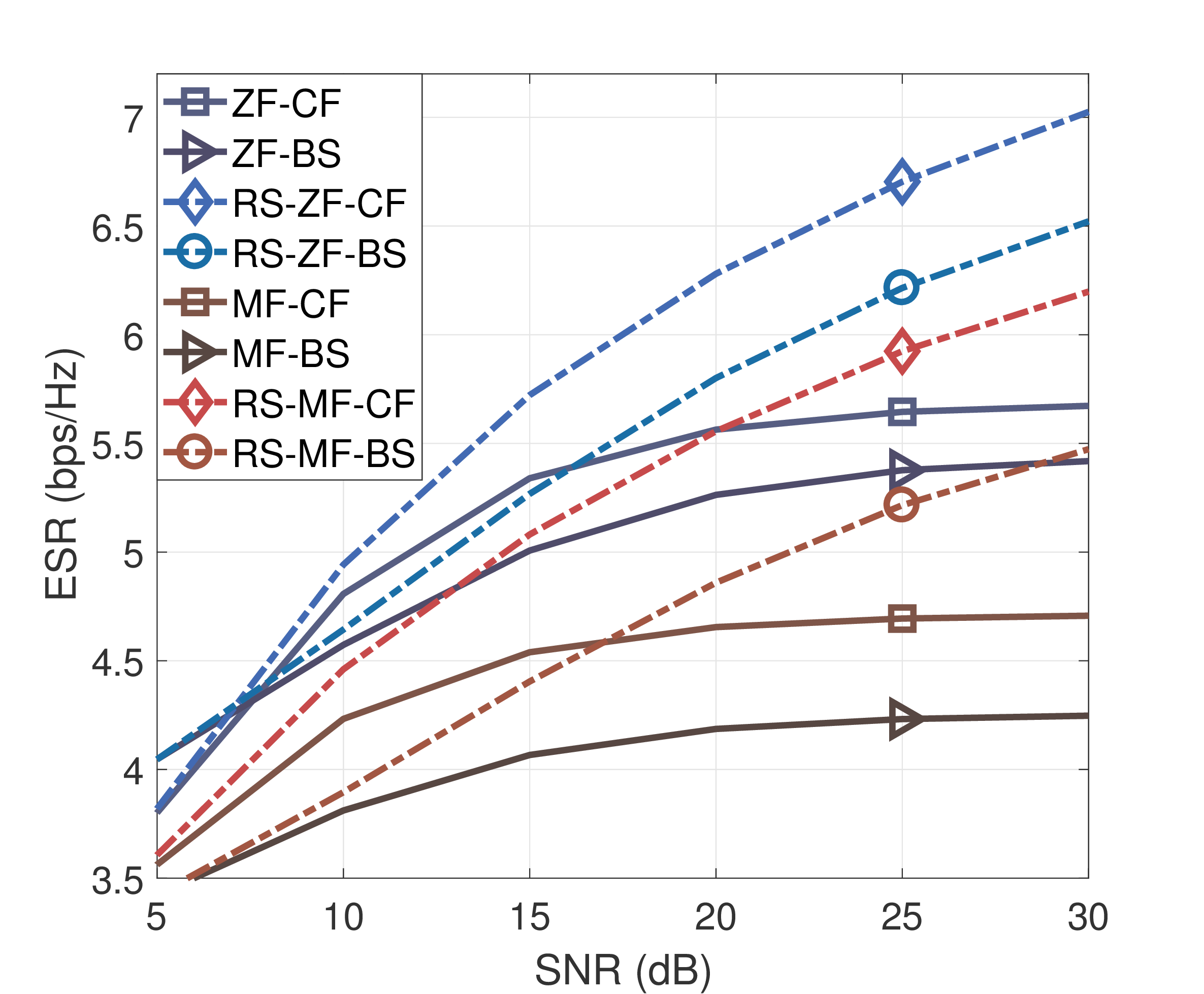}
\caption{Sum-rate performance of MIMO-BS and MIMO CF employing RS under imperfect CSIT, $M=6$, $K=3$, $\sigma_{e}^2=0.25$.} 
\label{FigC1}
\end{center}
\end{figure}
 We assessed the performance of the proposed RS-CF configuration through numerical experiments. We modeled the large-scale fading coefficients as
 \begin{equation}
     \zeta_{m,k}=P_{m,k}\cdot 10^{\frac{\sigma^{\left(\textrm{s}\right)}z_{m,k}}{10}},
 \end{equation}
 where $P_{m,k}$ represents the path loss and the parameter $10^{\frac{\sigma^{\left(\textrm{s}\right)}z_{m,k}}{10}}$ models the shadowing effects with standard deviation $\sigma^{\left(\textrm{s}\right)}=8$ dB and the random variable $z_{m,k}$ is Gaussian distributed with zero mean and unit variance. The path loss is calculated using a three-slope model as\par\noindent\small
 \begin{align}
     P_{m,k}=\begin{cases}
  -L-35\log_{10}\left(d_{m,k}\right), & \text{$d_{m,k}>d_1$} \\
  -L-15\log_{10}\left(d_1\right)-20\log_{10}\left(d_{m,k}\right), & \text{$d_0< d_{m,k}\leq d_1$}\\
    -L-15\log_{10}\left(d_1\right)-20\log_{10}\left(d_0\right), & \text{otherwise,}
\end{cases}
 \end{align}\normalsize
 where $d_{m,k}$ is the distance between the $m$-th AP and $k$-th users, $d_1=50$ m, $d_0= 10$ m, and the attenuation $L$ is given by\par\noindent\small
 \begin{align}
     L=&46.3+33.9\log_{10}\left(f\right)-13.82\log_{10}\left(h_{\textrm{AP}}\right)\nonumber\\
     &-\left(1.1\log_{10}\left(f\right)-0.7\right)h_u+\left(1.56\log_{10}\left(f\right)-0.8\right),
 \end{align}\normalsize
 where $h_{\textrm{AP}}=15$ m and $h_{u}=1.65$ are the positions of, respectively, AP and user equipment above the ground and frequency $f= 1900$ MHz.
 
 The noise variance is 
 \begin{equation}
     \sigma_w^2=T_o k_B B N_f,
 \end{equation}
 where $T_o=290$ K is the noise temperature, $k_B=1.381\times 10^{-23}$ J/K is the Boltzmann constant, $B=20$ MHz is the bandwidth and $N_f=9$ dB is the noise figure. The SNR is \par\noindent\small
 \begin{equation}
     \text{SNR}=\frac{P_{t}\textrm{Tr}\left(\mathbf{G}^{\text{T}}\mathbf{G}^{*}\right)}{M K \sigma_w^2}.
 \end{equation}\normalsize
 
\begin{figure}[ht]
\begin{center}
\includegraphics[width=1.0\columnwidth]{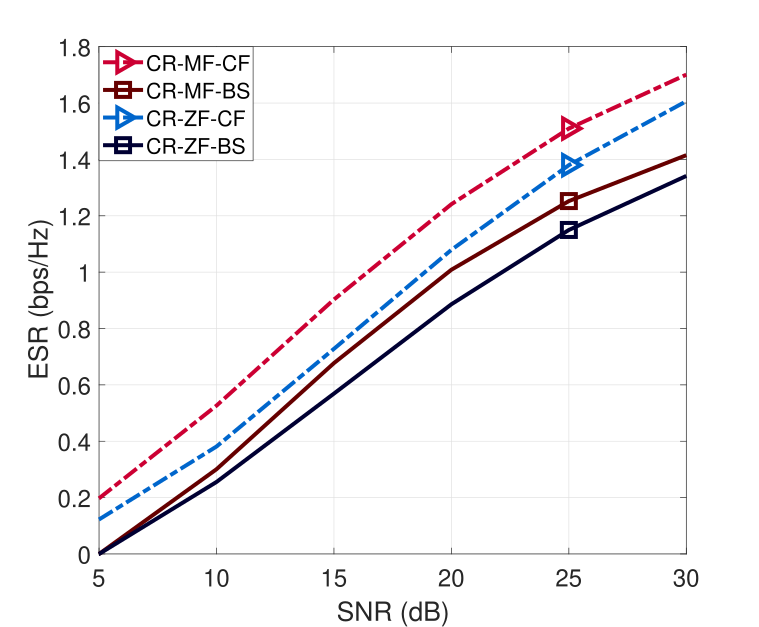}
\caption{CR obtained by the MIMO-BS and MIMO CF under imperfect CSIT, $M=6$, $K=3$, $\sigma_{e}^2=0.25$.} 
\label{FigC2}
\end{center}
\end{figure}
 We first considered the CF downlink in a cluster with $6$ single antenna APs, which are independently and uniformly distributed in a 600$\times$600 m$^2$ region, in which 3 users are placed uniformly at random. Fig.~\ref{FigC1} shows the performance of a standard MU-MIMO system with a BS placed at the center of the area. The power allocated to the common stream was found by employing exhaustive search with a grid step of $0.001$, whereas we employ uniform power allocation across private streams. The ASR 
 for every specific channel realization was computed using $100$ error matrices $\tilde{\mathbf{G}}$. On the other hand, a total of $100$ channel realizations were employed to obtain the ESR, i.e., resulting in a total of $10000$ trials. The results show a consistent gain, which is up to 10\%, of the proposed RS-CF over conventional schemes.

\begin{figure}[t]
\begin{center}
\includegraphics[width=1.0\columnwidth]{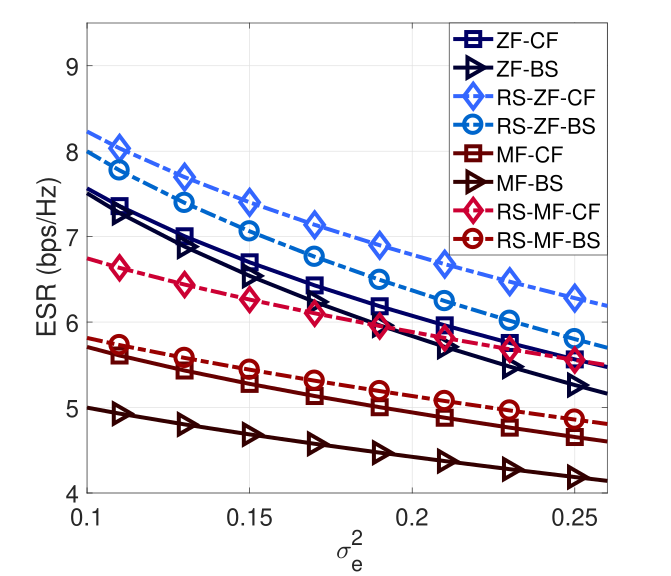}
\caption{ESR obtained by the MIMO-BS and MIMO CF under imperfect CSIT, $M=6$, $K=3$, SNR=$20$ dB.} 
\label{FigC3}
\end{center}
\end{figure}

 
 Fig.~\ref{FigC2} employs the same setup as in Fig.~\ref{FigC1} and shows the CR obtained by the systems using CF and BS. We observe that, in Fig. \ref{FigC2}, the CR increases with the transmit power. Next, we investigated the ESR performance with the increase in the variance of the error in the channel estimate. We set the SNR to $20$ dB and the other parameters were kept the same as before. Fig. \ref{FigC3} shows that RS-CF has the best performance with a consistent gain of approximately $1$ dB across all error variances. 

\section{Conclusions}
\label{sec:summ}
We proposed a novel wireless communications architecture that combines the benefits of CF and RS techniques to yield a network that is more robust to imperfect CSIT. In particular, we employed RS with linear precoders and derived closed-form expressions to compute the sum-rate performance of the proposed RS-CF system. Numerical results demonstrated that our RS-CF system with MF and ZF precoders outperform existing RS-BS-MF, RS-BS-ZF, and standard CF-MF, CF-ZF, BS-MF and BS-ZF systems   
in terms of sum-rate. When compared to RS-based systems, the RS-CF ESR shows up to $10$\% improvement.



\balance
\bibliographystyle{IEEEtran}
\bibliography{main}

\begin{thebibliography}{10}
\providecommand{\url}[1]{#1}
\csname url@samestyle\endcsname
\providecommand{\newblock}{\relax}
\providecommand{\bibinfo}[2]{#2}
\providecommand{\BIBentrySTDinterwordspacing}{\spaceskip=0pt\relax}
\providecommand{\BIBentryALTinterwordstretchfactor}{4}
\providecommand{\BIBentryALTinterwordspacing}{\spaceskip=\fontdimen2\font plus
\BIBentryALTinterwordstretchfactor\fontdimen3\font minus
  \fontdimen4\font\relax}
\providecommand{\BIBforeignlanguage}[2]{{%
\expandafter\ifx\csname l@#1\endcsname\relax
\typeout{** WARNING: IEEEtran.bst: No hyphenation pattern has been}%
\typeout{** loaded for the language `#1'. Using the pattern for}%
\typeout{** the default language instead.}%
\else
\language=\csname l@#1\endcsname
\fi
#2}}
\providecommand{\BIBdecl}{\relax}
\BIBdecl

\bibitem{Elhoushy2021}
S.~Elhoushy, M.~Ibrahim, and W.~Hamouda, ``Cell-free massive mimo: A survey,''
  \emph{IEEE Communications Surveys Tutorials}, Oct. 2021.

\bibitem{Ngo2017}
H.~Q. Ngo, A.~Ashikhmin, H.~Yang, E.~G. Larsson, and T.~L. Marzetta,
  ``Cell-free massive {MIMO} versus small cells,'' \emph{IEEE Transactions on
  Wireless Communications}, vol.~16, no.~3, pp. 1834--1850, 2017.

\bibitem{Nguyen2017}
L.~D. Nguyen, T.~Q. Duong, H.~Q. Ngo, and K.~Tourki, ``Energy efficiency in
  cell-free massive {MIMO} with zero-forcing precoding design,'' \emph{IEEE
  Communications Letters}, vol.~21, no.~8, pp. 1871--1874, 2017.

\bibitem{Tse2005}
D.~Tse and P.~Viswanath, \emph{Fundamentals of Wireless Communication},
  Cambridge, Ed.\hskip 1em plus 0.5em minus 0.4em\relax Cambridge University
  Press, 2005.

\bibitem{Nayebi2017}
E.~Nayebi, A.~Ashikhmin, T.~L. Marzetta, H.~Yang, and B.~D. Rao, ``Precoding
  and power optimization in cell-free massive {MIMO} systems,'' \emph{IEEE
  Transactions on Wireless Communications}, vol.~16, no.~7, pp. 4445--4459,
  2017.

\bibitem{Bjoernson2020}
E.~Björnson and L.~Sanguinetti, ``Scalable cell-free massive mimo systems,''
  \emph{IEEE Transactions on Communications}, vol.~68, no.~7, pp. 4247--4261,
  2020.

\bibitem{Palhares2021a}
V.~M.~T. Palhares, R.~C. de~Lamare, A.~R. Flores, and L.~T.~N. Landau,
  ``Iterative {AP} selection, {MMSE} precoding and power allocation in
  cell-free massive {MIMO} systems,'' \emph{IET Communications}, vol.~14,
  no.~22, pp. 3996--4006, 2021.

\bibitem{Vu2007}
M.~Vu and A.~Paulraj, ``{MIMO} wireless linear precoding,'' \emph{IEEE Signal
  Processing Magazine}, vol.~24, no.~5, pp. 86--105, Sep. 2007.

\bibitem{Clerckx2016}
B.~{Clerckx}, H.~{Joudeh}, C.~{Hao}, M.~{Dai}, and B.~{Rassouli}, ``Rate
  splitting for {MIMO} wireless networks: {A} promising {PHY}-layer strategy
  for {LTE} evolution,'' \emph{IEEE Communications Magazine}, vol.~54, no.~5,
  pp. 98--105, 2016.

\bibitem{Mao2018}
Y.~{Mao}, ``Rate-splitting multiple access for downlink communications
  systems,'' Ph.D. dissertation, The University of Hong Kong, 2018.

\bibitem{Mao2020}
Y.~{Mao} and B.~{Clerckx}, ``Beyond dirty paper coding for multi-antenna
  broadcast channel with partial {CSIT}: {A} rate-splitting approach,''
  \emph{IEEE Transactions on Communications}, vol.~68, no.~11, pp. 6775--6791,
  2020.

\bibitem{Clerckx2020}
B.~{Clerckx}, Y.~{Mao}, R.~{Schober}, and H.~V. {Poor}, ``Rate-splitting
  unifying {SDMA}, {OMA}, {NOMA}, and multicasting in {MISO} broadcast channel:
  {A} simple two-user rate analysis,'' \emph{IEEE Wireless Communications
  Letters}, vol.~9, no.~3, pp. 349--353, 2020.

\bibitem{Naser2020}
S.~{Naser}, P.~C. {Sofotasios}, L.~{Bariah}, W.~{Jaafar}, S.~{Muhaidat},
  M.~{Al-Qutayri}, and O.~A. {Dobre}, ``Rate-splitting multiple access:
  {U}nifying {NOMA} and {SDMA} in {MISO} {VLC} channels,'' \emph{IEEE Open
  Journal of Vehicular Technology}, vol.~1, pp. 393--413, 2020.

\bibitem{Jaafar2020}
W.~Jaafar, S.~Naser, S.~Muhaidat, P.~C. Sofotasios, and H.~Yanikomeroglu,
  ``Multiple access in aerial networks: {F}rom orthogonal and non-orthogonal to
  rate-splitting,'' \emph{IEEE Open Journal of Vehicular Technology}, vol.~1,
  pp. 372--392, 2020.

\bibitem{Han1981}
T.~S. {Han} and K.~{Kobayashi}, ``A new achievable rate region for the
  interference channel,'' \emph{IEEE Transactions on Information Theory},
  vol.~27, no.~1, pp. 49--60, 1981.

\bibitem{Carleial1978}
A.~{Carleial}, ``Interference channels,'' \emph{IEEE Transactions on
  Information Theory}, vol.~24, no.~1, pp. 60--70, 1978.

\bibitem{Haghi2021}
A.~Haghi and A.~K. Khandani, ``Rate splitting and successive decoding for
  {G}aussian interference channels,'' \emph{IEEE Transactions on Information
  Theory}, vol.~67, no.~3, pp. 1699--1731, 2021.

\bibitem{Yang2013}
S.~{Yang}, M.~{Kobayashi}, D.~{Gesbert}, and X.~{Yi}, ``Degrees of freedom of
  time correlated {MISO} broadcast channel with delayed {CSIT},'' \emph{IEEE
  Transactions on Information Theory}, vol.~59, no.~1, pp. 315--328, 2013.

\bibitem{Piovano2017}
E.~{Piovano} and B.~{Clerckx}, ``Optimal {DoF} region of the {$K$}-user {MISO
  BC} with partial {CSIT},'' \emph{IEEE Communications Letters}, vol.~21,
  no.~11, pp. 2368--2371, 2017.

\bibitem{JoudehClerckx2016}
H.~{Joudeh} and B.~{Clerckx}, ``Sum-rate maximization for linearly precoded
  downlink multiuser {MISO} systems with partial {CSIT}: A rate-splitting
  approach,'' \emph{IEEE Transactions on Communications}, vol.~64, no.~11, pp.
  4847--4861, Nov. 2016.

\bibitem{Hao2015}
C.~{Hao}, Y.~{Wu}, and B.~{Clerckx}, ``Rate analysis of two-receiver {MISO}
  broadcast channel with finite rate feedback: {A} rate-splitting approach,''
  \emph{IEEE Transactions on Communications}, vol.~63, no.~9, pp. 3232--3246,
  2015.

\bibitem{Joudeh2017}
H.~{Joudeh} and B.~{Clerckx}, ``Rate-splitting for max-min fair multigroup
  multicast beamforming in overloaded systems,'' \emph{IEEE Transactions on
  Wireless Communications}, vol.~16, no.~11, pp. 7276--7289, 2017.

\bibitem{Lu2018}
G.~{Lu}, L.~{Li}, H.~{Tian}, and F.~{Qian}, ``{MMSE}-based precoding for rate
  splitting systems with finite feedback,'' \emph{IEEE Communications Letters},
  vol.~22, no.~3, pp. 642--645, 2018.

\bibitem{Flores2020}
A.~R. {Flores}, R.~C. {De Lamare}, and B.~{Clerckx}, ``Linear precoding and
  stream combining for rate splitting in multiuser {MIMO} systems,'' \emph{IEEE
  Communications Letters}, vol.~24, no.~4, pp. 890--894, 2020.

\bibitem{rsthp}
A.~R. Flores, R.~C. De~Lamare, and B.~Clerckx, ``{T}omlinson-{H}arashima
  precoded rate-splitting with stream combiners for {MU-MIMO} systems,''
  \emph{IEEE Transactions on Communications}, 2021, in press.

\bibitem{Li2020}
Z.~Li, C.~Ye, Y.~Cui, S.~Yang, and S.~Shamai, ``Rate splitting for
  multi-antenna downlink: {P}recoder design and practical implementation,''
  \emph{IEEE Journal on Selected Areas in Communications}, vol.~38, no.~8, pp.
  1910--1924, 2020.

\bibitem{Joham2005}
M.~Joham, W.~Utschick, and J.~A. Nossek, ``Linear transmit processing in {MIMO}
  communications systems,'' \emph{IEEE Transactions on Signal Processing},
  vol.~53, no.~8, pp. 2700--2712, Aug. 2005.

\bibitem{Tuy2000}
H.~{Tuy}, ``Monotonic optimization: Problems and solution approaches,''
  \emph{SIAM Journal on Optimization}, vol.~11, no.~2, pp. 464--494, Feb. 2000.

\end{thebibliography}

\end{document}